\documentclass[12pt,english]{article}

\usepackage[latin9]{inputenc}
\usepackage{geometry}
\usepackage{youngtab}

\usepackage{bbm}
\usepackage{babel}
\usepackage{cite}
\usepackage[T1]{fontenc}
\usepackage{comment}
\usepackage{verbatim}
\usepackage{prettyref}
\usepackage{amsmath}
\usepackage{amssymb}
\usepackage{mathabx}
\usepackage{mathrsfs}
\usepackage{graphicx}
\usepackage{esint}
\usepackage{physics}
\usepackage{slashed}
\usepackage{subcaption}
\usepackage{array}   
\newcolumntype{C}{>{$}c<{$}}
\usepackage{dsfont}
\usepackage{xcolor}
\usepackage{color}
\definecolor{darkgreen}{rgb}{0,0.5,0}
\definecolor{darkblue}{rgb}{0,0,0.6}
\definecolor{purple2}{rgb}{0.4,.2,0.7}
\usepackage[colorlinks=true,citecolor=purple,linkcolor=purple2,urlcolor=purple2]{hyperref}

\definecolor{editgreen}{rgb}{0.0,0.55,0.1}

\definecolor{newblue}{rgb}{0.0,0.1,0.8}

\numberwithin{equation}{section}
\numberwithin{figure}{section}
\numberwithin{table}{section}

\newcommand{\rme}{\mathrm e}

\addtocounter{page}{-2}

\begin{document}

\begin{center}
{\LARGE A m\'enage \`a quatre without strings attached}

    \vspace{1.5cm}
    {\large Iosif Bena$^1$, Gongrui Cheng$^{2}$, Klaas Parmentier$^3$,\\ 
    Dimitrios Toulikas$^4$, Thomas Van Riet$^{2}$}\\
    \vspace{1cm}

    {$^1$\emph{Institut de Physique Th\'eorique, Universit\'e Paris Saclay,
    CEA,\\ CNRS, F-91191 Gif-sur-Yvette, France.}\\
    $^2$\emph{Instituut voor Theoretische Fysica, K.U. Leuven,\\
        Celestijnenlaan 200D B-3001 Leuven, Belgium.\\}
    $^3$\emph{Department of Applied Mathematics and Theoretical Physics,\\
University of Cambridge, Cambridge CB3 0WA, UK.}\\
    $^4$\emph{Department of Physics, Ben-Gurion University of the Negev,\\
     Beer Sheva 84105, Israel.}
    }
    
    \vspace{2cm}
    \end{center}

When two D-branes with eight mixed Dirichlet-Neumann directions cross, a fundamental string is created. This is the Hanany-Witten (HW) effect. In supergravity, the string shows up as an electric NSNS B-field sourced by the RR fields of the two branes. We show that two brane pairs, each of which would create an HW string on its own, can combine into a quadruple bound state with no HW strings attached. We explicitly construct a background with the fluxes of such a D0-D8-D4-D4' quadruple. Requiring no HW strings imposes a quadratic relation among the fluxes. We also realize this bound state as a supersymmetric solution of the non-Abelian DBI action of the D8-branes, whose Killing spinors match those of the bulk supergravity solution. The fluxed D8 stack acts as a domain wall between two supersymmetric D0-D8-D4-D4' backgrounds. Each background satisfies the no-string constraint, and the jumps in the fluxes give the charges carried by the stack.

\thispagestyle{empty}

\setcounter{tocdepth}{2}

\newpage
\thispagestyle{empty}
\tableofcontents
\newpage

\section{Introduction}
Intersecting branes in string theory can preserve a fraction of the supersymmetry. The rules for determining the orientations of these branes can be derived either in the full string theory, the brane worldvolume theories, or by using the supergravity solutions sourced by the branes~\cite{Douglas:1995bn, Gauntlett:1996pb, Tseytlin:1996bh, Bergshoeff:1996rn, Hanany:1996ie, Argurio:1997gt,Bena:2009fi}.
 
The best known brane intersections preserving 8 supercharges are  Dp-Dp' intersections that have 4 directions with mixed Dirichlet/Neumann boundary conditions (ND$=4$), as well as the U-duals thereof: Dp-NS5 intersections in which the Dp brane is along all but one of the NS5 directions, F1 strings orthogonal to Dp branes, or F1 strings parallel to NS5 branes. Supersymmetry requires ND$=0,4$ or $8$. The ND$=8$ class is the more subtle one and can exist instead as a triple bound state preserving 8 supercharges. 
However, this brane triplet is not egalitarian. One brane forms ND$=4$ pairs (or U-duals thereof) with each of the other two members of the triplet. But these two members in turn form a pair with 8 Neumann-Dirichlet directions (or U-dual thereof). Furthermore, when these branes cross they cannot go ``around'' each other, and they always produce the third kind of brane, by the Hanany-Witten effect \cite{Hanany:1996ie}. 

To illustrate this we can focus on the D0-D8-F1 system  \cite{Danielsson:1997wq, Witten:2000mf, Janssen:1999sa, Massar:1999sb, Imamura:2001cr, Bergshoeff:2003sy} : 
\begin{center}
\begin{tabular}{ c || c | c  c c c c c c c c}
D0 & $\star$ & - & - & - & - & - &  -  & - & -  &-  \\
D8 & $\star$ & $\star$ & $\star$ & $\star$ & $\star$ & $\star$ & $\star$ & $\star$ & $\star$  & -   \\
F1 & $\star$ & - & - & - & - & - &  -  & - & -  & $\star$ \\
\end{tabular}
\end{center}
The F1 string is orthogonal to the D0 branes and to the D8 branes, so the D0-F1 and the D8-F1 pairs are dual to an ND=4 pair of D branes. However, the D0 branes and the D8 branes have eight ND directions. Since D8 branes are spacetime domain walls, D0 branes cannot go around them, and when they pierce a D8 brane they create an F1 string by the Hanany-Witten effect. 

If one T-dualizes this system along four of the D8 brane directions, one obtains a D4-D4'-F1 configuration. When the D4 and D4' branes cross they also create an F1 string.

\begin{center}
\begin{tabular}{ c || c | c  c c c c c c c c}
D4 & $\star$ & $\star$ & $\star$ & $\star$ & $\star$ & - &  -  & - & -  &-  \\
D4' & $\star$ & - & - & - & - & $\star$ & $\star$ & $\star$ & $\star$  & -   \\
F1 & $\star$ & - & - & - & - & - &  -  & - & -  & $\star$ \\
\end{tabular}
\end{center}

The purpose of this paper is to show that a quadruple bound state (instead of a triplet) of branes, as depicted in Table \ref{table:D0D8D4}, gives a mechanism to not have any strings left over, essentially because the strings in both diagrams cancel out. In Section \ref{section2} we discuss how Hanany-Witten brane creation is visible in Supergravity. In Section \ref{sec:DBI} we present a non-Abelian DBI construction of a stack of D8 branes with worldvolume fluxes that give rise to D4, D4' and D0 charges. In Section \ref{section4} we construct the supergravity solutions corresponding to these branes, and show that they preserve the same supersymmetries as the DBI solution. In Section  \ref{sec:tensions} we use a superpotential analysis to determine when a given D8-D4-D4'-D0 brane bound state can be a domain wall between two supergravity solutions with these charges. We present our conclusions in Section \ref{section6}.

\begin{table}[h!]
\centering
\begin{tabular}{ c || c | c  c c c c c c c c}
D0 & $\star$ & - & - & - & - & - &  -  & - & -  &-  \\
D8 & $\star$ & $\star$ & $\star$ & $\star$ & $\star$ & $\star$ & $\star$ & $\star$ & $\star$  & -   \\
D4 & $\star$ & $\star$ & $\star$ & $\star$ & $\star$ & - &  -  & - & -  &-  \\
D4' & $\star$ & - & - & - & - & $\star$ & $\star$ & $\star$ & $\star$  & -   \\
\end{tabular}
\caption{D0-D8-D4-D4' bound state.}
\label{table:D0D8D4}
\end{table}


\section{Brane creation effects from supergravity}
\label{section2}
The equations of supergravity provide a quick way to understand Hanany-Witten (HW)-type brane creation. Consider the equation for the self-dual $F_5$ in Type-IIB supergravity:
\begin{equation}
    dF_5 = H_3 \wedge F_3 +\ldots \,,
\end{equation}
where the dots represent delta-function source terms. Using the fact that NS5 branes are magnetic sources for $H_3$ and D5 branes are magnetic sources for $F_3$, it is clear that a configuration with NS5 branes along (012345) and D5 branes along (012789) will have a non-zero $H_3 \wedge F_3$, and hence a nontrivial $F_5$. This is exactly the $F_5$ that is sourced by the D3 branes created by the Hanany-Witten effect when the D5 and NS5 branes cross each other.

The Hanany-Witten creation of F1 strings when the D0 and D8 branes cross each other can be understood similarly. The $B_2$ equation of motion is
\begin{equation}\label{eq:Beom}
d (e^{-2\phi }\star H_3)  = F_0 F_8 + \frac{1}{2}F_4\wedge F_4 - F_2\wedge F_6\,,
\end{equation}
where the Hodge star is defined using the string-frame metric. 
A D0 brane and a D8 brane are magnetic sources for $F_8$ and $F_0$, which in turn create a source for the $B$-field, which can be thought of as sourced by the HW string. A similar argument can be made for the D4-D4' system, which can be obtained by T-dualizing 4 times the D8-D0 one. The magnetic $F_4$ fluxes sourced by the D4 and D4' branes source a nontrivial $B_2$, corresponding to the HW string.

However, having both D8-D0 and D4-D4' fluxes present gives the option to cancel the terms on the right hand side against each other. This predicts brane bound states without an HW string attached, on the condition that the bulk fluxes are tuned such that:
\begin{equation}\label{Bianchi1}
     F_0 F_8 + \frac{1}{2}F_4\wedge F_4 =0\,.
\end{equation}
This local constraint must hold separately in each bulk region between branes.

Note that one can also try to achieve this with the $F_2\wedge F_6$ term, coming from intersecting D2 and D6 branes, which also create an F1 string by the HW effect:
\begin{center}
\begin{tabular}{ c || c | c  c c c c c c c c}
F1 & $\star$ & - & - & - & - & - &  -  & - & -  &$\star$  \\
D2 & $\star$ & $\star$ & $\star$ & - & - & - &  -  & - & -  &-  \\
D6 & $\star$ & - & - & $\star$ & $\star$ & $\star$ & $\star$ & $\star$ & $\star$  & -   \\
\end{tabular}
\end{center}

In this paper we set these D2-D6 charges to zero, as they increase complexity without bringing any new physics. 

In the next section we begin our analysis of the D8-D4-D4'-D0 system using the DBI action of D8 branes and, in the following sections we describe this system in supergravity. 

\section{The D8-D4-D4'-D0 solution of the nonabelian DBI action}
\label{sec:DBI}
For a stack of $N$ coincident Dp-branes with worldvolume coordinates $\sigma^a$, the non-abelian DBI and WZ actions are given by \cite{Myers1999, Tseytlin1997}:
\begin{align}
    S_{\text{DBI}}&=-\frac{T_p}{g_s} \int d^{p+1}\sigma\,\text{STr}\left(\sqrt{-\text{det}\left(
P\left[E_{ab}+E_{ai}(Q^{-1}-\delta)^{ij}E_{jb}\right]+
\lambda\,F_{ab}\right)\,\det(Q^i{}_j)}
\right),
\label{DBI} \\
S_{\text{WZ}}&=-T_p\int \text{STr}\left(P\left[e^{i \lambda\,i_\Phi i_\Phi}\left(\sum C^{(n)}e^B \right)\right]
e^{\lambda\,F}\right) \,,
\label{WZ}
\end{align}
where $E_{\mu \nu}=g_{\mu \nu}+B_{\mu \nu}$, $Q^i_{\,j} = \delta^i_{\,j}+i \lambda [ \Phi^i , \Phi^k] E_{kj}$, $\lambda= 2\pi \ell_s^2$ and STr denotes the symmetrized trace over gauge indices. This prescription is not exact for general non-commuting field strengths, with corrections appearing at order $F^6$ \cite{Denef2000, Sevrin2001}. However, for the constant, mutually commuting fluxes we use in subsection \ref{sec:generalDBI}, it reduces to a sum of Abelian DBI actions, avoiding such ambiguities.

Let us now consider the action of a stack of D8 branes in flat space. In the static gauge, $(x^a=\sigma^a, x^9=\lambda \Phi) $, the worldvolume action simplifies to:
\begin{align}
    S_{\text{DBI}}&=-\frac{T_8}{g_s} \int d^9x \, \text{STr} \left( \sqrt{-\text{det} \left( \eta_{ab}+\lambda F_{ab} +\lambda^2 D_a \Phi D_b \Phi \right)} \right) \,,\label{DBI action} \\
    S_{\text{WZ}}&=-T_8 \int \, \text{STr} \left(
P[C^{(9)}]
+\lambda\,P[C^{(7)}]\wedge F
+\frac{\lambda^2}{2}\,P[C^{(5)}]\wedge F\wedge F \right. \nonumber \\
& \quad \quad \quad \quad \quad \quad \left. + \frac{\lambda^3}{3!}\,P[C^{(3)}]\wedge F^3
+\frac{\lambda^4}{4!}\,P[C^{(1)}]\wedge F^4
\right) \,. \label{DBI Action:WZ}
\end{align}

From the WZ action we can see that the lower-brane charges inside the D8 are encoded by $\text{Tr}(F),\text{Tr}(F^2),\text{Tr}(F^3)$, and $\text{Tr}(F^4)$. We are interested in a configuration on the D8 worldvolume spanned by $(012345678)$ that has no D2 and D6 charges, and has dissolved D0 charge as well as dissolved D4 charges along directions $(01234)$ and $(05678)$.

\subsection{Fluxed D8 stacks with D0, D4 and D4' charges}\label{sec:generalDBI}

Consider $N$ coincident D8 branes along $(012345678)$ with a worldvolume flux that is block-diagonal in the four 2-planes $(12)$, $(34)$, $(56)$ and $(78)$:
\begin{equation}\label{eq:Fblocks}
    F=\sum_{k=1}^4 F_k\, dx^{2k-1}\wedge dx^{2k}\,,\qquad [F_k,F_l]=0\,,
\end{equation}
with $F_k$ commuting Hermitian $N\times N$ matrices, which we may simultaneously diagonalize. Since $F_k$ commute, the symmetrized trace in \eqref{DBI action} reduces to an ordinary trace. The WZ couplings  $\text{Tr}(F)$, $\text{Tr}(F^2)$ and $\text{Tr}(F^3)$ in general induce D6 charges, D4 charges along all six 4-planes, and D2 charges. We are interested in stacks that carry only the charges of Table \ref{table:D0D8D4}. This requires
\begin{equation}\label{eq:zerotrace}
    \text{Tr}\,F_k=0 \,, \quad  \text{Tr}\,(F_k F_l)=0 \; \text{for}\; (k,l) \notin \{(1,2),(3,4)\} \,, \quad  \text{Tr}\,(F_k F_l F_m)=0 \,.
\end{equation}
It is not hard to see that these conditions imply that the stack must contain at least four D8-branes.
The remaining charges are
\begin{equation}\label{eq:generalcharges0}
    q_8=N\,,\qquad q_{4'}=\text{Tr}(F_1 F_2)\,,\qquad q_4=\text{Tr}(F_3F_4)\,,\qquad q_0=\text{Tr}(F_1 F_2F_3F_4)\,.
\end{equation}
The DBI action treats the four 2-planes democratically and it is the absence of mixed D4 charges in \eqref{eq:zerotrace} that groups them into the two 4-planes $(1234)$ and $(5678)$.

The $\kappa$-symmetry analysis of the next subsection shows that, once \eqref{eq:zerotrace} is imposed, such a configuration is supersymmetric only if the fluxes in the two 2-planes of each 4-plane are equal up to a sign, $F_2=\eta_1F_1$ and $F_4=\eta_2F_3$ with $\eta_{1,2}=\pm1$. Thus the D8 worldvolume flux must be self-dual or anti-self-dual in each 4-plane. Within this class, the most general supersymmetric ansatz is therefore
\begin{equation}\label{eq:generalF}
    F=A\left(dx^1\wedge dx^2+\eta_1\,dx^3\wedge dx^4\right)+B\left(dx^5\wedge dx^6+\eta_2\,dx^7\wedge dx^8\right)\,,\qquad [A,B]=0\,.
\end{equation}
For this flux, the trace conditions \eqref{eq:zerotrace} collapse to
\begin{equation}\label{eq:traceconditions}
    \text{Tr}A=\text{Tr}B=\text{Tr}(AB)=\text{Tr}(A^2B)=\text{Tr}(AB^2)=0\,,
\end{equation}
and the charges \eqref{eq:generalcharges0} become\footnote{We suppress the overall WZ normalisations and the appropriate powers of $\lambda$; with $u_i=\lambda^2a_i^2$ and $v_i=\lambda^2 b_i^2$ for the eigenvalues $a_i$, $b_i$ of $A$ and $B$, the charges read $q_{4'}=\eta_1\sum_i u_i$, $q_4=\eta_2\sum_iv_i$ and $q_0=\eta_1\eta_2\sum_iu_iv_i$. Note that the D4$'$ along $(05678)$ is induced by the flux in the $(1234)$ plane and vice versa.}
\begin{equation}\label{eq:generalcharges}
    q_8=N\,,\qquad q_{4'}=\eta_1\,\text{Tr}A^2\,,\qquad q_4=\eta_2\,\text{Tr}B^2\,,\qquad q_0=\eta_1\eta_2\,\text{Tr}(A^2B^2)\,.
\end{equation}
In particular,
\begin{equation} \label{eq:signs}
    \text{sign}(q_0q_8) = \text{sign}(q_4q_{4'}) = \eta_1\eta_2\,.
\end{equation}

\subsection{DBI Killing spinor analysis from \texorpdfstring{$\kappa$}{}-symmetry}\label{sec:kappa}

We now determine the supersymmetry preserved by the fluxed D8 stack from $\kappa$-symmetry, in a flat background with $B_2=0$. We write the IIA supersymmetry parameter as a doublet $\epsilon=(\epsilon_1,\epsilon_2)^T$ of Majorana--Weyl spinors of opposite chirality and set ${\cal F}\equiv\lambda F$.  For a Dp-brane with $p$ even the preserved supersymmetries satisfy $(1+\Gamma_\kappa)\epsilon=0$ with
\begin{equation}\label{eq:Gammakappa}
    \Gamma_\kappa=\frac{\sqrt{-\det g}}{\sqrt{-\det(g+{\cal F})}}\sum_{n=0}^{\infty}\frac{1}{2^nn!}\,{\cal F}_{i_1j_1}\cdots{\cal F}_{i_nj_n}\gamma^{i_1j_1\cdots i_nj_n}\,\sigma_3^{\,n+p/2}\,i\sigma_2\,\Gamma_{(0)}\,,
\end{equation}
where $\Gamma_{(0)}$ is the standard product of induced gamma matrices along the worldvolume (see for example ~\cite{Bergshoeff:1996tu, Cederwall:1996ri}). The $\Gamma$ matrices act on the spinor index of $\epsilon_{1,2}$, while the Pauli matrices $\sigma_i$ act on the doublet $(\epsilon_1,\epsilon_2)$. It is convenient to define the constant 
%
%
involutions associated with the four brane species in Table \ref{table:D0D8D4},
\begin{equation}\label{eq:Pdef}
    P_8=i\sigma_2\Gamma^{012345678}\,,\quad P_0=i\sigma_2\Gamma^{0}\,,\quad P_4=i\sigma_2\Gamma^{01234}\,,\quad P_{4'}=i\sigma_2\Gamma^{05678}\,,
\end{equation}
and the four mutually commuting operators $J_1=\Gamma^{12}\sigma_3$, $J_2=\Gamma^{34}\sigma_3$, $J_3=\Gamma^{56}\sigma_3$, $J_4=\Gamma^{78}\sigma_3$, which square to $-1$ and anticommute with $P_8$.

Consider now the flux ansatz \eqref{eq:Fblocks}. Because the $F_k$ commute, we can simultaneously diagonalize them,
\begin{equation}\label{eq:Fabel}
{\cal F}^{(I)}=\sum_k\lambda f^{(I)}_k dx^{2k-1}\wedge dx^{2k}\,,
\end{equation}
and no ordering prescription is needed to apply \eqref{eq:Gammakappa} to the stack. The $\kappa$-matrix of the stack is also diagonal in the gauge indices: $\Gamma_\kappa=\text{diag}(\Gamma_{\kappa,1},\ldots,\Gamma_{\kappa,N})$, where $I=1,\ldots,N$ labels the D8-branes in the stack and $\Gamma_{\kappa,I}$ is the Abelian expression \eqref{eq:Gammakappa} for the $I$-th brane. Given the Abelian field strengths \eqref{eq:Fabel} the sum in \eqref{eq:Gammakappa} factorizes (see appendix \ref{app:kappafact}) into:
\begin{equation}\label{eq:GammakappaI}
    \Gamma_{\kappa,I}=\prod_{k=1}^4\frac{\mathbbm{1}+\lambda f^{(I)}_kJ_k}{\sqrt{1+\lambda^2(f^{(I)}_k)^2}}\;P_8=e^{X_I}P_8\,,\qquad X_I=\sum_{k=1}^4\theta_k^{(I)}J_k\,,\quad \tan\theta_k^{(I)}=\lambda f_k^{(I)}\,,
\end{equation}
where we used $J_k^2=-1$ to exponentiate each factor. Since $P_8X_I=-X_IP_8$, each $\Gamma_{\kappa,I}$ squares to one and each brane in the stack separately preserves 16 supercharges. 
The combined stack then preserves the common solutions of $(1+\Gamma_{\kappa,I})\epsilon=0$ for all $I$. As shown in Appendix \ref{app:kappacommon}, once the conditions \eqref{eq:zerotrace} are imposed such common solutions exist only if
\begin{equation}\label{eq:thetapairing}
    \theta^{(I)}_2=\eta_1\theta^{(I)}_1\,,\qquad \theta^{(I)}_4=\eta_2\theta^{(I)}_3\,,
\end{equation}
with the same $\eta_{1,2}=\pm1$ for every brane in the stack. This is the origin of the ansatz \eqref{eq:generalF}: supersymmetry forces $|f_1|=|f_2|$ and $|f_3|=|f_4|$ brane by brane. Altogether, for \eqref{eq:generalF} with arbitrary commuting $A$ and $B$, the preserved supersymmetries are exactly those obeying the three constant, mutually commuting projection conditions
\begin{equation}\label{eq:threeproj}
    P_8\epsilon=-\epsilon\,,\qquad \Gamma^{1234}\epsilon=\eta_1\epsilon\,,\qquad \Gamma^{5678}\epsilon=\eta_2\epsilon\,.
\end{equation}
The bound state thus preserves 4 of the 32 supercharges, independent of the flux values.\footnote{The counting is the same as that of a flat D0-D8-D4-D4$'$ intersection: 32 supercharges are halved by each of the three independent projectors.}

Using $\Gamma^{1234}P_8=P_{4'}$, $\Gamma^{5678}P_8=P_4$ and $\Gamma^{12345678}P_8=P_0$, the projection conditions \eqref{eq:threeproj} become
\begin{equation}\label{eq:fourproj}
    P_8\epsilon=-\epsilon\,,\qquad P_{4'}\epsilon=-\eta_1\epsilon\,,\qquad P_4\epsilon=-\eta_2\epsilon\,,\qquad P_0\epsilon=-\eta_1\eta_2\epsilon\,,
\end{equation}
which are the projector conditions of a D8, a D4$'$, a D4 and a D0 
which satisfy the identity $P_8=P_0P_4P_{4'}$. The signs in \eqref{eq:fourproj} are precisely the signs of the induced charges \eqref{eq:generalcharges}: the D4$'$ charge has the sign of $\eta_1$, the D4 charge the sign of $\eta_2$ and the D0 charge the sign of $\eta_1\eta_2$. The identity $P_0P_8=P_4P_{4'}$ among the brane involutions is therefore the spinorial counterpart of the sign relation \eqref{eq:signs}: a supersymmetric D8 stack can carry a D0-D8 pair and a D4-D4$'$ pair of either sign, but $q_0q_8$ and $q_4q_{4'}$ always have the same sign. In the branch where all four charges are positive, $\eta_1=\eta_2=+1$, the flux is self-dual in both 4-planes and
\begin{equation}\label{eq:allpositiveproj}
  P_0\epsilon=P_4\epsilon=P_{4'}\epsilon=P_8\epsilon=-\epsilon\,.
\end{equation}
In the next section we describe this system in supergravity and recover this Killing spinor condition from the supergravity Killing spinor equations in section \ref{sec:susy}.

\section{Supergravity solutions}
\label{section4}
\subsection{The Ansatz}
To describe the supergravity solution sourced by the branes in Table \ref{table:D0D8D4}, we begin with the Einstein-frame 10-dimensional supergravity ansatz:
\begin{align}
ds^2 =&-e^{2B(y)+2C(y)}dt^2  + e^{2A(y)}dy^2  \nonumber\\
&+e^{2B(y)-\frac{C(y)}{4}}\left( e^{2D(y)}\left(d\theta_1^2+\ldots + d\theta_4^2\right) + e^{-2D(y)}\left(d\theta_5^2+\ldots + d\theta_8^2\right) \right)  \,, \label{ansatz}
\end{align}
where $A(y)$ is pure gauge. Aside from the dilaton $\phi(y)$ we have the following field ans\"atze:
\begin{align}
F_0& =Q_8\,,\\
F_8 &=Q_0 d\theta^1\wedge\ldots d\theta^8 \,,\\
F_4 &=Q_4 d\theta^5\wedge\ldots d\theta^8 + Q_{4'}d\theta^1\wedge\ldots d\theta^4\,,
\end{align}
with the RR fluxes given in their magnetic description. As usual, $F_0$ is constant (up to jumps). 

In the action and in the supersymmetry variations, we use the electric representative of D0 charge, defined in our conventions by \footnote{Note $F_8 = -e^{\frac{3}{2}\phi} \star_{10} F_2$. We choose the ten-dimensional orientation
$\text{vol}_{10} = e^{A+9B} dt\wedge dy  \wedge d\theta^1\wedge\ldots d\theta^8$.
}
\begin{equation}
    F_2 = Q_0 e^{A-7B+2C-\frac{3}{2}\phi} dt \wedge dy\,.
\end{equation}

Before continuing, note how the $B_2$ equation of motion immediately constrains the bulk fluxes: since $H_3=0$ in our ansatz (no Hanany-Witten strings), we must obey \eqref{Bianchi1}, which, given the fluxes above, implies
\begin{equation}\label{eq:bulkconstraint}
Q_0Q_8 + Q_4Q_{4'} = 0\,,
\end{equation}
\emph{everywhere in the bulk}, since \eqref{Bianchi1} is a point-wise condition and the fluxes are piecewise constant. Note that \eqref{eq:bulkconstraint} constrains the bulk fluxes both on the right and on the left of the D8 stack, but not the charges, $q_i$, carried by this stack. These charges are equal to the jumps of the bulk fluxes across the wall: $q_i = Q_i^R-Q_i^L$. Note also that Equation \eqref{eq:bulkconstraint} forces $Q_4Q_{4'}<0$ in any region where $Q_0Q_8>0$. In the rest of this paper we will choose an orientation in which  $Q_0,Q_8,Q_4>0$ and $Q_{4'}<0$.
 
\subsection{Worldvolume tadpole and no-string condition}\label{sec:tadpole}

If one considers a D8-D4-D4'-D0 brane stack (of section \ref{sec:generalDBI}) that separates two bulk regions, the local constraint \eqref{eq:bulkconstraint} applies to the fluxes in each region separately. We now relate this bulk constraint to the worldvolume gauge-field tadpole.


Even if this codimension-one stack causes a jump in the fluxes that extends to infinity, we can work in a limit where the charges of the stack are much smaller than those of the background, and describe this D8 stack using its DBI and WZ actions as in section \ref{sec:generalDBI}. We then vary the WZ action with respect to the overall $U(1)$ worldvolume gauge field. Using $\delta\,\text{STr}F^k=k\,d\,\text{STr}(\delta A\wedge F^{k-1})$ and integrating by parts, the exterior derivative on the RR potentials then produces the bulk field strengths $F_n=dC_{n-1}$.\footnote{The massive completion of the D8 WZ action, which contains the Chern--Simons term $-\frac{\lambda^5F_0}{5!}\int\omega_9(A)$ with $d\omega_9=\text{Tr}F^5$, supplies the $F_0$ term.} One finds $\delta S_{WZ}=-\lambda T_8\int J\wedge\delta A$ with the tadpole 8-form
\begin{equation}\label{eq:tadpoleform}
    J=F_8\,\text{Tr}\mathds 1+F_6\wedge\lambda\,\text{Tr}F+F_4\wedge\frac{\lambda^2}{2}\text{Tr}F^2+F_2\wedge\frac{\lambda^3}{3!}\text{Tr}F^3+F_0\,\frac{\lambda^4}{4!}\text{Tr}F^4\,.
\end{equation}
This is the source term in the worldvolume Gauss law and measures the number of fundamental strings ending on the stack. Each term pairs a bulk flux with the dissolved charge of the brane that has eight ND directions with respect to it: the bulk D0 flux with the D8 charge ($q_8Q_0$), the bulk D4 flux with the dissolved D4$'$ charge ($q_{4'}Q_4$), the bulk D4$'$ flux with the dissolved D4 charge ($q_4Q_{4'}$) and the Romans mass with the dissolved D0 charge ($q_0Q_8$). These are exactly the pairings that create Hanany--Witten strings, in agreement with \eqref{eq:Beom}; the D2-D6 pairing sits in the $F_6\wedge\text{Tr}F$ term that we have switched off.

Since the bulk fluxes jump across the stack, they must be evaluated with some care. Regulating the wall as a slab of thickness $\epsilon$ with uniform charge density $q_i/\epsilon$, the Bianchi identities make each bulk flux linear inside the slab, $F_0(y)=Q_8^L+\frac{q_8}{\epsilon}(y-y_0)$ etc., and the integrated tadpole is $\int_{y_0}^{y_0+\epsilon}dy\,\frac{q_0}{\epsilon}F_0(y)=q_0\bar Q_8$, independent of $\epsilon$. The fluxes therefore enter through their average $\bar{Q}_i= \frac{1}{2} (Q_i^L+Q_i^R)$. Demanding that no string ends on the stack gives
\begin{equation}\label{eq:tadpole}
    {\cal T}\equiv q_0\bar Q_8+q_8\bar Q_0+q_4\bar Q_{4'}+q_{4'}\bar Q_4=0\,.
\end{equation}
Defining ${\cal C}(Q)\equiv Q_0Q_8+Q_4Q_{4'}$ and using the jump condition $q_i=Q_i^R-Q_i^L$, one can check that
\begin{equation}\label{eq:CRCL}
    {\cal C}(Q^R)-{\cal C}(Q^L)={\cal T}
\end{equation}
identically.


This equation is an important self-consistency condition. It indicates that the domain walls that separate regions with no F1 charges will not source F1 charge. Hence, setting to zero the NS-NS 3-form field strength is consistent both in the bulk and on the brane.



\subsection{BPS equations from the superpotential}
The second-order equations of motion can be derived from a one-dimensional effective action $S=\int dy (T- V)$, where $T$ is the kinetic term
\begin{equation}
T =-\frac{1}{2}G_{IJ}\dot{\Phi}^I\dot{\Phi}^J\,,
\end{equation}
and $\Phi^I = \{\phi, B, C, D\}$. The one-dimensional action follows by reducing the following truncation of the ten-dimensional action:
\begin{equation}\label{eq:lag}
    \mathcal{L}= \sqrt{-g}\left(\mathcal{R}-\tfrac12(\partial\phi)^2- \frac{1}{48}\rme^{\phi/2}F_{\mu\nu\rho\sigma}F^{\mu\nu\rho\sigma} - \frac{1}{4}\rme^{3\phi/2}F_{\mu\nu}F^{\mu\nu}- \frac12 M^2 \rme^{5\phi/2}\right)\,.
\end{equation}
The first two terms in the action contribute the kinetic term:
\begin{equation}
-\frac{1}{2}G_{IJ}\dot{\Phi}^I\dot{\Phi}^J = e^{9B-A}\left(-\frac{1}{2}\dot{\phi}^2 + 72 \dot{B}^2  - \frac{9}{8}\dot{C}^2 -8\dot{D}^2\right)      \,.
\end{equation}
This shows that the easiest gauge choice (corresponding to fixing the $y$ coordinate) is the one where $A=9B$, such that the effective action is
\begin{equation}
\int dy\left(-\frac{1}{2}\dot{\phi}^2 + 72 \dot{B}^2  - \frac{9}{8}\dot{C}^2 -8\dot{D}^2 - V\right) \,.
\end{equation}
This is supplemented with the ``zero-energy'' constraint, $T+V = 0$, from varying the action with respect to the $A$-field.  The potential $V$ comes from reducing the RR-flux terms in 10 dimensions:
\begin{equation}
V = \frac{1}{2}e^{A+9B}\left(Q_8^2e^{\frac{5}{2}\phi} + Q_4^2 e^{-8B + C +8D +\frac{1}{2}\phi}+Q_{4'}^2 e^{-8B + C -8D+\frac{1}{2}\phi} + Q_0^2e^{-16B +2C -\frac{3}{2}\phi}    \right)\,.
\end{equation}

As usual, one expects supersymmetry to simplify the resulting second-order equations to first-order ones, in the form of a gradient flow derived from a superpotential $W$. This superpotential $W$ should obey the Hamilton-Jacobi equation,
$V =\frac{1}{2}G^{IJ}\partial_IW\partial_JW$, and the corresponding flow equations automatically satisfy the zero energy constraint:
\begin{equation}
\dot{\Phi}^I=\pm G^{IJ}\partial_JW\,.
\end{equation}
Crucially, for flows corresponding to BPS brane bound states, one expects $W$ to be a sum of terms, one for each brane species. For our quadruple bound state, this would mean
\begin{equation}\label{eq:Wsum}
W= W_8 + W_4 + W_{4'} + W_0\,,
\end{equation}
where each term is of the form $W_k=Q_k\tau_k(\Phi)$, with $Q_k$ the bulk flux and $\tau_k(\Phi)$ the DBI tension factor (the Einstein-frame volume element times the appropriate dilaton power) of the corresponding brane. The $W_k$ themselves are not brane tensions: as we explain in section~\ref{sec:tensions}, the tension of a brane located at $y_0$ is the \emph{jump} $\Delta W_k=q_k\tau_k(\Phi(y_0))$ of the corresponding term across $y_0$, where $q_k=Q_k^R-Q_k^L$ is the charge of the brane. The fact that $W$ is simply a sum of such terms reflects the absence of binding energies. Therefore, prior to a Killing-spinor analysis we simply write down the sum of the expected superpotential terms from the ansatz and Table \ref{table:D0D8D4}:
\begin{equation}\label{eq:W}
W =  Q_8 e^{9B +\frac{5}{4}\phi} + Q_4 e^{5B + \frac{C}{2} + 4D +\frac{1}{4}\phi} + Q_{4'}e^{5B + \frac{C}{2} - 4D +\frac{1}{4}\phi} +Q_{0}e^{B+C-\frac{3}{4}\phi} \,,
\end{equation}
with the individual superpotential terms in \eqref{eq:Wsum} corresponding to
\begin{align}\label{eq:Widef}
W_8 &= Q_8 e^{9B+\frac54\phi}\,, \qquad
W_0 = Q_0 e^{B+C-\frac34\phi}\,, \nonumber\\
W_4 &= Q_4 e^{5B+\frac12 C+4D+\frac14\phi}\, , \qquad
W_{4'} = Q_{4'} e^{5B+\frac12 C-4D+\frac14\phi}\, .
\end{align}
We then find
\begin{equation}\label{eq:susycorss}
\frac{1}{2}G^{IJ}\partial_I W\partial_JW = V - (Q_0Q_8 + Q_{4}Q_{4'})e^{10B +C +\frac{1}{2}\phi}\,.
\end{equation}
Remarkably, the bulk flux constraint \eqref{eq:bulkconstraint} ensures the Hamilton-Jacobi equation is exactly satisfied.
The resulting first-order equations are
\begin{align}
 \dot{\phi} &= \frac{5}{4}W_8 + \frac{1}{4}\left(W_4 + W_{4'} \right)  -\frac{3}{4}W_0\,, \label{eq:dotphi}\\
 \dot{B} &= -\frac{1}{16}W_8 - \frac{5}{144}\left(W_4 + W_{4'} \right)  -\frac{1}{144}W_0\,,\\
 \dot{C} &= \frac{2}{9}\left(W_4 + W_{4'} \right)  +\frac{4}{9}W_0\,,\\
 \dot{D} &=  \frac{1}{4}\left(W_4 - W_{4'} \right) \label{eq:dotD}\,.
\end{align}
We now demonstrate that these can indeed be recovered from a Killing-spinor analysis.

\subsection{Supersymmetry analysis}\label{sec:susy}
We now plug the ansatz \ref{ansatz} in the 10-dimensional supersymmetry variations. The dilatino variation is~\cite{VanPol:2015yok}\footnote{Our definition of $F_2$ differs by an overall sign from that used in \cite{VanPol:2015yok}.}
\begin{equation}\label{eq:dilatino}
    \delta \lambda = -\frac12 \slashed\partial\phi\, \epsilon- \frac58 M \rme^{5\phi/4}\,\epsilon+\frac{3}{16}\rme^{3\phi/4} \slashed{F}^{(2)}\Gamma_{11}\,\epsilon - \frac{1}{192}\rme^{\phi/4} \slashed{F}^{(4)}\,\epsilon  \,,
\end{equation}
while the gravitino variation takes the form \cite{VanPol:2015yok}
\begin{equation}\label{eq:gravitino}\begin{split}
      \delta \psi_\mu = &\left(\partial_\mu  + \frac14 \omega_\mu^{ab}\Gamma_{ab}\right)\epsilon - \frac{1}{32}M \rme^{5\phi/4} \Gamma_\mu \epsilon-\frac{1}{64}\rme^{3\phi/4}\left(\Gamma_\mu \slashed{F}^{(2)}-16 F_{\mu\rho}\Gamma^{\rho}\right)\Gamma_{11}\epsilon\\
      &+\frac{1}{256}\rme^{\phi/4}\left(\Gamma_\mu \slashed{F}^{4}+\frac{32}{3}\Gamma^{\rho \sigma\nu}F_{\rho\sigma\nu\mu}\right)\epsilon \,.
      \end{split}
\end{equation}
We will impose the following projection conditions:
\begin{equation}\label{eq:proj}
    \Gamma^{\hat 1 \hat 2 \hat 3 \hat 4} \epsilon = \Gamma^{\hat 5 \hat 6 \hat 7 \hat 8} \epsilon = - \Gamma^{\hat y}\epsilon = \epsilon\,.
\end{equation}
These are the same projection conditions as those in \eqref{eq:allpositiveproj} found from $\kappa$-symmetry. The first two are simply identical. The third is the D8 projection condition $P_8\epsilon=-\epsilon$: in the 32-component notation of \cite{VanPol:2015yok} the doublet operator $i\sigma_2\Gamma^{\hat t\hat 1\cdots\hat 8}$ becomes, up to a convention-dependent sign, $\Gamma^{\hat t\hat 1\cdots \hat 8}\Gamma_{11}\,\propto\;\Gamma^{\hat y}$. Then $P_8\epsilon=-\epsilon$ is the transverse projection condition $\Gamma^{\hat y}\epsilon=-\epsilon$ of a D8 domain wall, and the D0, D4 and D4$'$ projection conditions follow from $P_8=P_0P_4P_{4'}$ exactly as in \eqref{eq:fourproj}. 

    Setting the dilatino variation \eqref{eq:dilatino} to zero  gives:
\begin{align}
    \delta \lambda ={}&
    \Bigl(
    \frac 1 2 e^{-A} \phi '
    -\frac 5 8 Q_8 e^{\frac 5 4 \phi}
    +\frac 3 8 Q_0 e^{-8B+C-\frac3 4 \phi}
    \nonumber\\
    & -\frac 1 8 Q_{4'}e^{-4B+\frac C 2 -4D +\frac 1 4 \phi}
    -\frac 1 8 Q_{4}e^{-4B+\frac C 2 +4D +\frac 1 4 \phi}
    \Bigr) \epsilon=0 \,,
\end{align}
which, in the gauge \(A=9B\), is precisely \eqref{eq:dotphi}.
Similarly, the $t$-component of \eqref{eq:gravitino} gives:
\begin{equation}
    B'+C'=-\frac{1}{16}W_8+\frac {7}{16} W_0 +\frac {3}{16} W_{4'} +\frac {3}{16} W_{4}\,,
\end{equation}
while the $\theta_{1234}$ and $\theta_{5678}$ components of \eqref{eq:gravitino} give:
\begin{equation}
    B'-\frac{C'}{8}+D' = -\frac{1}{16}W_8 -\frac {1}{16} W_0 +\frac {3}{16} W_{4} -\frac {5}{16} W_{4'} \, ,
\end{equation}
\begin{equation}
    B'-\frac{C'}{8}-D' = -\frac{1}{16}W_8-\frac {1}{16} W_0 +\frac {3}{16} W_{4'} -\frac {5}{16} W_{4}\, .
\end{equation}
Altogether, this reproduces the first-order equations  \eqref{eq:dotphi}-\eqref{eq:dotD}.

\subsection{The solution}\label{sec:solution}
The equations above can be solved explicitly. With our choice of signs, $Q_0Q_8 >0$ and $Q_4Q_{4'}<0$, the D0 and D8 charges appear  naturally in hyperbolic functions, while the D4 and D4' charges appear  naturally in trigonometric functions:
\begin{align}
e^B =& \,(\text{sinh}\,X)^{\frac{5}{72}}\,(\text{sin}\,2Y)^{\frac{5}{144}}\,\bigg(\sqrt{\frac{Q_0}{Q_8}}\,\text{coth}\,X\bigg)^{\frac{1}{144}}\,, \label{eq:B}\\
e^C =& \,(\text{sinh}\,X)^{-\frac{4}{9}}\,(\text{sin}\,2Y)^{-\frac{2}{9}}\,\bigg(\sqrt{\frac{Q_8}{Q_0}}\,\text{tanh}\,X\bigg)^{\frac{4}{9}}\,,\\
e^D =& \,\bigg(\sqrt{-\frac{Q_{4'}}{Q_4}}\,\text{tan}\,Y\bigg)^{\frac{1}{4}} \,,\\
e^{\phi} =&\,(\text{sinh}\,X)^{-\frac{1}{2}}\,(\text{sin}\,2Y)^{-\frac{1}{4}}\,\bigg(\sqrt{\frac{Q_0}{Q_8}}\,\text{coth}\,X\bigg)^{\frac{3}{4}}\,, \label{eq:phi}
\end{align}
with
\begin{equation}
    X = \sqrt{Q_8Q_0}\,(y_1-y)\,, \quad Y=\sqrt{-Q_{4'}Q_4}\,(y-y_2)\,,
\label{eq:y1y2}
\end{equation}
defined on the interval where $\sinh X$ and $\sin 2Y$ are positive: ~$y_2 < y < \text{min}\big(y_1,\, y_2 + \tfrac{\pi}{2}(-Q_4Q_{4'})^{-1/2}\big)$. On the flow one finds the simple expressions
\begin{align}
W_8 & = \sqrt{Q_0Q_8}\,\coth X\,,\quad\,\, W_0 = \sqrt{Q_0Q_8}\,\tanh X\,, \nonumber\\
 W_4 & = \sqrt{-Q_4Q_{4'}}\,\tan Y\,,\quad W_{4'} = -\sqrt{-Q_4Q_{4'}}\,\cot Y\,.
\end{align}
This configuration solves the first-order equations \eqref{eq:dotphi}-\eqref{eq:dotD} together with the zero-energy constraint $T+V=0$.

In string frame the solution takes a familiar form. Recall that a Dp-brane with harmonic function $Z_p$ has $ds^2=Z_p^{-1/2}ds^2_\parallel+Z_p^{1/2}ds^2_\perp$ and $e^{\phi}=Z_p^{(3-p)/4}$. Applying the harmonic superposition rule to the four branes of Table \ref{table:D0D8D4} gives
\begin{align}\label{eq:harmonic}
ds^2_{10}=&-(Z_0Z_8Z_4Z_{4'})^{-\frac12}dt^2+(Z_0Z_8Z_4Z_{4'})^{\frac12}dy^2 \nonumber\\
&+\left(\frac{Z_0Z_{4'}}{Z_8Z_4}\right)^{\frac12}(d\theta_1^2+\cdots+d\theta_4^2)+\left(\frac{Z_0Z_{4}}{Z_8Z_{4'}}\right)^{\frac12}(d\theta_5^2+\cdots+d\theta_8^2)\,,\\
e^{\phi}=&\,Z_0^{\frac34}Z_8^{-\frac54}Z_4^{-\frac14}Z_{4'}^{-\frac14}\,.
\end{align}
One can then check that \eqref{eq:B} - \eqref{eq:phi} are precisely of this form with\footnote{The normalization of the $Z$'s is fixed only up to the rescaling $(Z_8,Z_0,Z_4,Z_{4'})\to(cZ_8,cZ_0,Z_4/c,Z_{4'}/c)$, which leaves \eqref{eq:harmonic} invariant.}
\begin{align}\label{eq:Zs}
    Z_8&=\sinh X\,,& Z_0&=\sqrt{\tfrac{Q_0}{Q_8}}\,\cosh X\,,\nonumber\\
    Z_4&=\sqrt2\left(\tfrac{Q_4}{-Q_{4'}}\right)^{\frac14}\cos Y\,,& Z_{4'}&=\sqrt2\left(\tfrac{-Q_{4'}}{Q_4}\right)^{\frac14}\sin Y\,.
\end{align}
The ``harmonic'' functions are not harmonic, but they satisfy the simple first-order system
\begin{equation}\label{eq:Zflow}
    Z_8'=-Q_8\,Z_0\,,\qquad Z_0'=-Q_0\,Z_8\,,\qquad Z_4'=-Q_4\,Z_{4'}\,,\qquad Z_{4'}'=-Q_{4'}\,Z_4\,,
\end{equation}
which makes the physics transparent: each $Z_k$ is driven by its Hanany--Witten partner, the brane with which it has eight ND directions. When the partner flux vanishes, its function is constant and $Z_k$ is linear, as for a single D8-brane. Conversely, for a D0 in a region with Romans mass, $Z_0$ is no longer linear, which is the supergravity avatar of the D0 being affected by the D8. The bulk constraint $Q_0Q_8=-Q_4Q_{4'}$ guarantees that the two sectors oscillate with the same ``frequency'' $\sqrt{Q_0Q_8}$ in $y$. Moreover, on the solution the four superpotential terms are simply
\begin{equation}\label{eq:WZ}
    W_k=-\frac{Z_k'}{Z_k}=-\frac{d}{dy}\ln Z_k\,,\qquad k=0,8,4,4'\,,
\end{equation}
so that $W=-\frac{d}{dy}\ln(Z_0Z_8Z_4Z_{4'})$, with $y$ the coordinate of the gauge $A=9B$. Since the $Z_k$ are continuous across a brane, the jump $\Delta W_k=-\Delta Z_k'/Z_k$ of a superpotential term across a wall is the jump in the slope of the $Z_k$ divided by its value there. This is the quantity that will compute the brane tensions in section \ref{sec:tensions}.

This solution matches the timelike T-dual of the D($-$1)-D7-D3 bound state solution found in \cite{Reymond:2024mwe}. Note that the analytic continuations involved in timelike T-duality interchange trigonometric and hyperbolic functions, which has to be kept in mind when comparing with the solution given there.

\section{Brane tensions}\label{sec:tensions}
\subsection{D8 brane
}

Recall the local form of the D8 brane solution in the 10-dimensional Einstein frame:
\begin{align}
& ds^2= H^{1/8}ds^2_9 +H^{9/8}dy^2\,,\\
& e^{\phi} = H^{-5/4}\,,\\
& H'=-F_0\,.
\end{align}
We fix $F_0>0$. Then $H$ decreases with $y$ and, after a shift of $y$, we can write $H=F_0\,(y_1-y)$. The massive solution ends at $y=y_1$, where $H$ vanishes and the dilaton diverges. A D8 brane is a domain wall across which $F_0$ jumps. Placing it at $y_0<y_1$ and setting $F_0=0$ for $y<y_0$ gives
\begin{align}
    y<y_0:&\qquad H(y) = F_0\,(y_1-y_0)\,,\\
    y_0<y<y_1:&\qquad H(y) = F_0\,(y_1-y)\,,
\end{align}
i.e. flat space with constant dilaton on the left, glued with a kink at $y_0$ to the massive region on the right. The end-point $y_1$ is the same one that bounds the full solution \eqref{eq:y1y2}.

As we show in general in Section~\ref{sec:junction}, the energy density per unit coordinate volume localized on a wall is the jump of the superpotential across it. For the D8
\begin{equation}
    \mathcal T_8 = W_8(y_0^+) - W_8(y_0^-) = \frac{F_0}{H(y_0)}\,,
\end{equation}
where $W_8 = F_0 H^{-1}$ as in \eqref{eq:Widef}, since
$B=\tfrac1{16}\log H$ and $\phi=-\tfrac54\log H$. Dividing by the proper
worldvolume element $e^{9B(y_0)}$ gives the Einstein-frame physical tension
\begin{equation}
T_8 = e^{-9B(y_0)}\,\mathcal T_8 = g_s^{5/4}\,F_0\,,\qquad
g_s = e^{\phi(y_0)}\,,
\end{equation}
or $T_8^{(s)} = e^{-9\phi/4}\,T_8 = F_0/g_s$ in string frame, the standard
D8 tension (in units where a single D8 brane carries $F_0$-jump one).

\subsection{D0-D8-D4-D4' bound state: junction conditions}\label{sec:junction}

We now show that the fluxed D8 stack of section~\ref{sec:DBI} can act as a BPS domain wall between two bulk regions described in section~\ref{section4}. The junction conditions require DBI energy of the fluxed D8 stack to equal the jump of the superpotential across it, and this saturates a BPS bound whose equality conditions reproduce the $\kappa$-symmetry results of section~\ref{sec:kappa}. Concretely, we place the fluxed D8 stack of Section \ref{sec:DBI} at a position $y_w$ in the supergravity solution and ask under which conditions it can act as the domain wall between two bulk regions with fluxes $Q^L_k$ and $Q^R_k$.

The charges are fixed by the Bianchi identity: the WZ couplings of the stack source $dF_0=q_8\,\delta(y-y_w)dy$, $dF_8=q_0\,\delta(y-y_w)dy\wedge d\theta^{1\cdots8}$ and $dF_4=(q_4d\theta^{5678}+q_{4'}d\theta^{1234})\wedge\delta(y-y_w)dy$, where $q_k$ are the charges \eqref{eq:generalcharges0}, so that $q_k=Q_k^R-Q_k^L$. In the language of \eqref{eq:Zflow}, the $q_k$ are the jumps in the slopes $\Delta Z_k'=-q_k Z_{\bar k}(y_w)$ with $\bar k$ the Hanany--Witten partner brane (with 8 ND directions) of $k$. Finally, the trace conditions \eqref{eq:zerotrace} are the statement that no other bulk flux (D6, D2 or mixed D4) jumps across the wall, as required by the supergravity ansatz.

The tensions follow from the junction conditions for the scalars $\Phi^I=(\phi,B,C,D)$. In the gauge $A=9B$ the one-dimensional action in the presence of the wall is
\begin{equation}
    S=\int dy\left[-\tfrac12 G_{IJ}\dot\Phi^I\dot\Phi^J-V\big(\Phi;Q(y)\big)-{\cal T}_{\rm DBI}(\Phi)\,\delta(y-y_w)\right]+S_{RR}\,,
\end{equation}
with $V$ piecewise constant in its $Q$-dependence, ${\cal T}_{\rm DBI}(\Phi)$ the DBI energy density of the stack per unit coordinate volume, and $S_{RR}$ (which contains the WZ term) independent of $\Phi$ at fixed charges. Integrating the scalar equations across the wall then gives $G_{IJ}[\dot\Phi^J]=\partial_I{\cal T}_{\rm DBI}(\Phi_w)$, where $[X] \equiv X(y_w^+)-X(y_w^-)$ denotes the jump of $X$ across the wall and $\Phi_w \equiv \Phi(y_w)$. On each side the flow equations hold with their own charges, $\dot\Phi^I=G^{IJ}\partial_JW^{L,R}$ with $W^{L,R}=\sum_kQ^{L,R}_k\tau_k$, and $\Phi$ is continuous, so the four junction conditions are
\begin{equation}\label{eq:junction}
    \partial_I{\cal T}_{\rm DBI}(\Phi_w)=\partial_I\Delta W(\Phi_w)\,,\qquad \Delta W\equiv W^R-W^L=\sum_k q_k\tau_k\,,\qquad I=\phi,B,C,D\,.
\end{equation}
The remaining equation, obtained by varying with respect to $A$ before fixing the gauge, is the zero-energy constraint $\tfrac12 G_{IJ}\dot \Phi^I\dot \Phi^J=V$ on either side. Subtracting the two sides and using \eqref{eq:junction} gives $\langle\dot \Phi\rangle^I\partial_I\mathcal{T}_{\rm DBI}=[V]$, with
$\langle X\rangle=\tfrac12\big(X(y_w^+)+X(y_w^-)\big)$.This is the equation of motion for the position of the wall. On the flows one finds
$\langle\dot \Phi\rangle^I\partial_I\Delta W-[V]=-\big(\mathcal{C}(Q^R)-\mathcal{C}(Q^L)\big)\tau_0\tau_8$, so by \eqref{eq:CRCL} this condition is precisely the vanishing of the worldvolume tadpole \eqref{eq:tadpole}.

Equation 
\eqref{eq:junction}  fixes the field-space gradient  $\partial_I\mathcal{T}_{\rm DBI}$ at $\Phi_w$. Its value follows from an Euler relation. The tension factors $\tau_k=\rme^{n_k\cdot\Phi}$ have exponent vectors $n_8=(\tfrac54,9,0,0)$, $n_4=(\tfrac14,5,\tfrac12,4)$, $n_{4'}=(\tfrac14,5,\tfrac12,-4)$ and $n_0=(-\tfrac34,1,1,0)$ in the basis $(\phi,B,C,D)$. Since every constituent of the stack extends along $t$, each $\tau_k$ contains $\sqrt{-g_{tt}}=\rme^{B+C}$ exactly once. The direction $\xi=(0,\tfrac19,\tfrac89,0)$ shifts $B+C$ by one unit and leaves $\phi$, $D$ and the spatial scale factors $\rme^{2B-C/4\pm2D}$ unchanged, so that $n_k\cdot\xi=1$ for all four species.\footnote{The relation $n_0+n_8=n_4+n_{4'}$ makes the $n_k$ linearly dependent, so $\xi$ is unique only up to the direction $(1,-\tfrac{5}{36},\tfrac89,0)$, along which all $\tau_k$ are invariant. The same relation makes the cross terms in \eqref{eq:susycorss} combine into the single combination $(Q_0Q_8+Q_4Q_{4'})\rme^{10B+C+\frac12\phi}$.} Hence each $\tau_k$, and therefore $\Delta W$, has weight one along $\xi$: $\xi^I\partial_I\Delta W=\Delta W$.
The DBI energy of the stack, for any flux of the form \eqref{eq:Fblocks}, is
\begin{equation}\label{eq:TDBI}
    {\cal T}_{\rm DBI}=\sum_{i=1}^N\tau_8\sqrt{(1+x^{(i)}_1)(1+x^{(i)}_2)}\sqrt{(1+x^{(i)}_3)(1+x^{(i)}_4)}\,,
\end{equation}
with $x^{(i)}_{1,2}=\lambda^2(f^{(i)}_{1,2})^2\,\tau_{4'}/\tau_8$ and $x^{(i)}_{3,4}=\lambda^2(f^{(i)}_{3,4})^2\,\tau_4/\tau_8$. Here each 2-plane contributes a factor $\rme^{4a}+\lambda^2f^2$ to the string-frame determinant, and we used $\tau_{4'}/\tau_8=\rme^{-4a_1}$ and $\tau_4/\tau_8=\rme^{-4a_2}$, where $\rme^{2a_{1,2}}=\rme^{\frac{\phi}{2}+2B-\frac{C}{4}\pm 2D}$ are the string-frame scale factors of the 4-planes $(1234)$ and $(5678)$. The ratios $\tau_{4'}/\tau_8$ and $\tau_4/\tau_8$ have weight zero along $\xi$, so ${\cal T}_{\rm DBI}$ is $\tau_8$ times a weight-zero function and has weight one as well. Contracting \eqref{eq:junction} with $\xi^I$ and using $\xi^I\partial_If=f$ for a weight-one function then determines the values at $\Phi_w$
\begin{equation}\label{eq:TequalsDW}
    {\cal T}_{\rm DBI}(\Phi_w)=\Delta W(\Phi_w)=\sum_kq_k\tau_k(\Phi_w)\,,
\end{equation}
which identifies the DBI energy of the stack with the total jump of the superpotential across the wall.

Equation \eqref{eq:TequalsDW} is very constraining because of a BPS-type bound. For $x,x'\geq0$ one has $\sqrt{(1+x)(1+x')}\geq1+\sqrt{xx'}$ with equality iff $x=x'$. Applying this to each 4-plane in~\eqref{eq:TDBI}, using $\tau_8\sqrt{x_1x_2x_3x_4}=\lambda^4|f_1f_2f_3f_4|\,\tau_4\tau_{4'}/\tau_8=\lambda^4|f_1f_2f_3f_4|\,\tau_0$, where the last step follows from $n_0+n_8=n_4+n_{4'}$, summing over the branes and using the triangle inequality for the charges~\eqref{eq:generalcharges0}, we find
\begin{equation}\label{eq:bound}
    {\cal T}_{\rm DBI}\geq N\tau_8+|q_{4'}|\tau_{4'}+|q_4|\tau_4+|q_0|\tau_0\geq q_8\tau_8+q_{4'}\tau_{4'}+q_4\tau_4+q_0\tau_0=\Delta W\,.
\end{equation}
Since \eqref{eq:TequalsDW} indicates  that the bound is saturated at $\Phi_w$, every inequality in \eqref{eq:bound} must be an equality. The first requires $(f_1^{(i)})^2=(f_2^{(i)})^2$ and $(f_3^{(i)})^2=(f_4^{(i)})^2$ on every brane; the factors $\lambda^2\rme^{-4a_{1,2}}$ cancel, so this condition is background independent. The triangle inequalities together with $|q_k|=q_k$ require $(f_1f_2)^{(i)}\geqslant0$ and $(f_3f_4)^{(i)}\geqslant0$ on every brane. Together, $f_2^{(i)}=f_1^{(i)}$ and $f_4^{(i)}=f_3^{(i)}$: the flux is of the form \eqref{eq:generalF} with $\eta_1=\eta_2=+1$ on every brane, which is exactly the branch selected by the projections \eqref{eq:proj}, equivalently \eqref{eq:allpositiveproj}. We thus recover from the bulk both the pairing and the signs found from $\kappa$-symmetry, and in particular all charges of the domain wall are non-negative:\footnote{Upon choosing different relative signs in the superpotential, $W=W_8+\eta_2W_4+\eta_1W_{4'}+\eta_1\eta_2W_0$, the same argument gives $\eta_1q_{4'}\geq0$, $\eta_2q_4\geq0$, $\eta_1\eta_2q_0\geq0$ and the pairing $f_2=\eta_1f_1$, $f_4=\eta_2f_3$: the relative orientations of the worldvolume flux are the signs with which $W_4$ and $W_{4'}$ enter the bulk superpotential. An anti-D8 stack, $q_8=-N$, has the same DBI energy but $|q_8|\tau_8>q_8\tau_8$, so the bound can never be saturated: there are no anti-D8 walls in this background.}
\begin{equation}\label{eq:positivejumps}
    q_8=N>0\,,\qquad q_{4'}=\mathrm{Tr}A^2\geqslant0\,,\qquad q_4=\mathrm{Tr}B^2\geqslant0\,, \qquad q_0=\mathrm{Tr}(A^2B^2)\geqslant0\,.
\end{equation}


Moreover, with the pairing imposed one finds that ${\cal T}_{\rm DBI}=\Delta W$ term by term and identically in $\Phi$, so that all four junction conditions \eqref{eq:junction} hold for any $\Phi_w$: the wall can sit at any $y_w$. This is the finite-charge version of the no-force condition, and the position of the stack is a modulus. This identifies the jumps $\Delta W_k=q_k\tau_k$ with the individual brane tensions, as anticipated.


Note that a wall with vacuum on one side, $Q^L_k=0$, would have $q_k=Q^R_k$ and the bulk constraint \eqref{eq:bulkconstraint} on the right would force $q_4q_{4'}<0$. This would correspond to a negative jump in the superpotential, $\Delta W_{4'}<0$, which is excluded by \eqref{eq:positivejumps}. Hence a string-free bound state with all charges non-vanishing is necessarily a wall between two non-trivial bulk regions. A simple example of bulk fluxes, in the basis $(Q_0,Q_8,Q_4,Q_{4'})$, is:
\begin{equation}\label{eq:wallexample}
Q^L=\Big(0,\,0,\,0,\,-q_{4'}-\frac{q_0q_8}{q_4}\Big)\,,\qquad
Q^R=\Big(q_0,\,q_8,\,q_4,\,-\frac{q_0q_8}{q_4}\Big)\,.
\end{equation}
There also exist solutions where all fluxes are non-vanishing on both sides. For instance, the minimal stack $N=4$ with $\lambda A=\mathrm{diag}(1,1,-1,-1)$ and $\lambda B=\mathrm{diag}(1,-1,1,-1)$ satisfies \eqref{eq:traceconditions} and carries $(q_0,q_8,q_4,q_{4'})=(4,4,4,4)$. It can separate, for example
\begin{equation}
    Q^L=(2,\,11,\,1,\,-22)\,,\qquad Q^R=(6,\,15,\,5,\,-18)\,,
\label{eq:wallexample2}
\end{equation}
for which ${\cal C}(Q^L)=22-22=0$ and ${\cal C}(Q^R)=90-90=0$. Both regions have the signs chosen in section~\ref{section4} and are therefore in the family of solutions constructed in section~\ref{sec:solution}.

\section{Discussion}
\label{section6}
The main message of this paper is that brane bound states with eight mixed Neumann-Dirichlet directions do not necessarily feature Hanany-Witten (HW) strings or their duals. We have specifically demonstrated this for the D0-D8 bound state which by itself would require the introduction of an HW string. However, by superposing it with a  D4-D4' bound state with HW strings, the two HW strings effectively cancel if the bulk fluxes away from this D8-D4-D4'-D0 bound state satisfy: 
\begin{equation}\label{eq:bulknostring}
Q_0Q_8 + Q_4Q_{4'}=0\,.
\end{equation}

The charges of the brane stack, $q_k$, are equal to the jumps of the bulk fluxes, $Q_k$, across the stack. However, they obey a different constraint, 
\begin{equation}\label{eq:boundary-susy}
\text{sign}(q_0q_8)=\text{sign}(q_4q_{4'})\,,
\end{equation}
imposed by supersymmetry. Furthermore, the absence of the worldvolume gauge-field tadpole only demands that the combination $Q_0Q_8+Q_4Q_{4'}$ be the same on either side of the stack, which is automatic when the bulk no-string condition \eqref{eq:bulknostring} holds everywhere. 

As a consequence, a string-free D8-D4-D4'-D0 bound state with all charges non-vanishing can never interpolate between the vacuum and a non-trivial flux background\footnote{Unless one adds so-called OF1 planes, then explicit examples with no fluxes on one side can arise \cite{Cribiori:2026qfs}.}: it is a domain wall between two regions that both carry the fluxes of a string-free D0-D8-D4-D4$'$ configuration.

We have verified this explicitly in the open-string picture, where we considered a D8 stack with commuting worldvolume fluxes of the form \eqref{eq:Fblocks} that induce D4 and D0 charges (but no D2, D6 or crossed D4 charges). The $\kappa$-symmetry analysis showed that supersymmetry requires the flux to be (anti-)self-dual in each 4-plane, which then preserves four supercharges. We explicitly constructed the corresponding supergravity solution and
verified it preserves an eighth of the 10d supersymmetry. We furthermore computed the tensions of the separate branes in the bound state. The junction conditions imply that the DBI energy of the stack equals the jump of the superpotential. 

Several interesting questions and puzzles remain. The first is that the sign of the bulk fluxes, \eqref{eq:bulknostring} and the sign of brane charges \eqref{eq:boundary-susy} are opposite, despite the fact that the supersymmetries of the bulk and of the brane are compatible. This is not a convention-dependent statement, and underlies the fact that the solutions of the DBI action cannot interpolate between vacuum and the solutions we construct in Section \ref{sec:solution}. 

Another is the global structure of our solutions. A solution describing multiple parallel D8 branes has a harmonic function, $Z_8$, that has an upward-pointing kink at every D8 location, and can be thought of as a part of a global solution that has two $O8^-$ orientifold planes at the two ends of the universe. In our solutions, all four functions $Z_k$ have kinks at the domain walls, but they are no longer linear in the bulk \eqref{eq:Zflow}. It would be interesting to see if one can obtain a global solution describing multiple supergravity regions of the form \eqref{eq:harmonic} bounded by D8-D4-D4'-D0 domain walls. It would be fascinating to ascertain whether such a global solution can be realized with only $O8^-$ end-of-the world branes, or whether more complicated end-of-the world branes are needed.

Our solution can be thought of as  the timelike T-dual of the D($-$1)-D7-D3 bound state of \cite{Reymond:2024mwe} (see also \cite{toappear}).\footnote{See also \cite{Billo:2021xzh} for a related D($-$1)-D7 system with magnetic worldvolume flux. However, their open-string analysis did not address the induced D3-brane charges or their supergravity interpretation.} By constructing the T-dual Lorentzian bound state in this paper we could appeal to well-established techniques and confirm the suspicion of \cite{Aguilar-Gutierrez:2022kvk, Reymond:2024mwe} that there must be D3-brane creation in a D($-$1)-D7 bound state.  Our improved understanding of this bound state, or the dual D0-D8-D4-D4', potentially impacts the classification of instantons in 8d and monopoles in 9d YM theories \cite{Billo:2021xzh, Nekrasov:2017cih, Nekrasov:2018xsb}.

Finally, it would be interesting to expand our analysis to D8-D4-D4'-D0 solutions of higher cohomogeneity, where some of the D4 and D0 branes are localized along the D8 worldvolume. These solutions generalize the  D8-D0 solutions of \cite{Imamura:2001cr} and would depend on two or three variables. We expect the scaling limits of these solutions \cite{BDTW-scaling} to give rise to solutions with AdS$_2 \times S^3$ factors, which would be interesting to construct.

\subsection*{Acknowledgments}
 We thank Bert Janssen for a useful discussion. The work of GC and TVR is supported by the KU Leuven C1 grant ZKE7799C16/25/010. GC also acknowledges support from the China Scholarship Council. KP was supported by the STFC consolidated grant ST/T000694/1 and Simons Investigator award \#620869. DT is supported by an Azrieli Fellowship, by the German-Israeli grant ``Holography and the Swampland", by Carole and Marcus Weinstein through the BGU Presidential Faculty Recruitment Fund, by the VATAT Research Hub in the Field of quantum computing, and by the ERC Starting Grant 101117338, dSHologQI. 

\appendix

\section{Details of the \texorpdfstring{$\kappa$}{kappa}-symmetry analysis}\label{app:kappa}

Below, we collect some technical details behind the $\kappa$-symmetry analysis of section\,\ref{sec:kappa}.

\subsection{Factorisation of \texorpdfstring{$\Gamma_\kappa$}{Gamma kappa}}\label{app:kappafact}

Here we explain how \eqref{eq:Gammakappa} factorizes into \eqref{eq:GammakappaI}, given the abelian field strengths \eqref{eq:Fabel}. 

Firstly, note that for the $I$-th D8-brane in the stack, the field strength \eqref{eq:Fabel} has support on four separate two-planes. The determinant therefore factorizes: 
\begin{equation}
    -\det\bigl(g+\mathcal F^{(I)}\bigr) = \prod_{k=1}^{4} \left[1+\lambda^2\bigl(f_k^{(I)}\bigr)^2\right] \,,
\end{equation}
while $-\det g=1$ in static gauge.

The numerator of \eqref{eq:Gammakappa} has a similar structure. Since the gamma matrices are fully antisymmetrized, a nonzero term can contain at most one field-strength component from each two-plane. For a choice of $n$ distinct planes, the $n!$ orderings and $2^n$ orientations of the index pairs cancel the factor $1/(2^n n!)$. Moreover, for $p=8$, the Pauli-matrix factor is $\sigma_3^n$, so we may attach one of these $\sigma_3$ to each of the $\Gamma^{2k-1,2k}$, and hence each selected plane contributes $\lambda f_k^{(I)}J_k$.

Summing over all choices of planes gives
\begin{equation}
    \Gamma_{\kappa,I} = \prod_{k=1}^{4} \frac{1+\lambda f_k^{(I)}J_k}{\sqrt{1+\lambda^2(f_k^{(I)})^2}}\;P_8.
\end{equation}
with the D8 involution as defined in the main text, $P_8= i\sigma_2\Gamma_{(0)}=i\sigma_2\Gamma^{012345678}$. 

The exponential form follows upon introducing the angles
\begin{equation}
    \theta_k^{(I)} = \arctan\!\bigl(\lambda f_k^{(I)}\bigr) \in \left(-\frac{\pi}{2},\frac{\pi}{2}\right) \,.
\end{equation}
Since $J_k^2=-1$, each factor can be written as
\begin{equation}
    \frac{1+\lambda f_k^{(I)}J_k} {\sqrt{1+\lambda^2(f_k^{(I)})^2}} = \cos\theta_k^{(I)}+J_k\sin\theta_k^{(I)} = e^{\theta_k^{(I)}J_k} \,.
\end{equation}
The $J_k$ commute, and hence
\begin{equation}
    \Gamma_{\kappa,I}=e^{X_I}P_8 \,, \qquad X_I=\sum_{k=1}^{4}\theta_k^{(I)}J_k \,,
    \label{eq:app-kappa-exponential}
\end{equation}
as stated in \eqref{eq:GammakappaI}.

\subsection{Common Killing spinors of the stack}\label{app:kappacommon}
A supersymmetry preserved by the stack must satisfy
\begin{equation}
    \Gamma_{\kappa,I}\epsilon=-\epsilon
\end{equation}
for every brane. Comparing two branes and using $P_8X_I=-X_IP_8$, we obtain
\begin{equation}
    \Gamma_{\kappa,J}\Gamma_{\kappa,I}\epsilon = e^{X_J-X_I}\epsilon = \epsilon \,.
    \label{eq:app-relative-kappa}
\end{equation}
Since the $J_k$ commute and square to $-1$, we may diagonalise them on the complexified spinor space. On each joint eigenspace, $J_k\epsilon=i\alpha_k\epsilon$, where $\alpha_k=\pm1$. The flux-dependent factor $e^{X_I}$ found in Appendix~\ref{app:kappafact} acts by multiplication by $e^{i\sum_k\alpha_k\theta_k^{(I)}}$. Equation \eqref{eq:app-relative-kappa} requires this phase to be the same for every brane:
\begin{equation}
    \sum_k\alpha_k\theta_k^{(I)} = c \pmod{2\pi} \,,
    \label{eq:app-common-phase}
\end{equation}
with $c$ independent of $I$. This relation does not yet require the angles to cancel in pairs. The additional restriction comes from the trace conditions \eqref{eq:zerotrace}.

We use units with $\lambda=1$ in the following. To relate the angle condition to the trace conditions, it is useful to express it in terms of products of fluxes. The identity 
\begin{equation}
    1+i\alpha_k f_k^{(I)} = \sec\theta_k^{(I)}\, e^{i\alpha_k\theta_k^{(I)}}
\end{equation}
is the eigenvalue version of the factorization in Appendix~\ref{app:kappafact}. The product of the four factors therefore obeys
\begin{equation}
    \prod_k\left(1+i\alpha_k f_k^{(I)}\right) = D_I e^{ic} \,, \qquad D_I \equiv \prod_k\sqrt{1+\bigl(f_k^{(I)}\bigr)^2}\,.
    \label{eq:app-flux-product}
\end{equation}
Its phase is fixed by \eqref{eq:app-common-phase}, while its expansion contains precisely the flux products that enter the trace conditions. Here $D_I$ is the DBI energy density of the $I$-th brane, with the common tension factor omitted.

After summation over the branes, the trace conditions \eqref{eq:zerotrace} remove the linear, cubic and mixed-quadratic terms on the left-hand side of \eqref{eq:app-flux-product}. Using the charges defined in \eqref{eq:generalcharges0}, this leaves 
\begin{equation}
    \mathcal E e^{ic} = N-\alpha_1\alpha_2q_{4'} -\alpha_3\alpha_4q_4+\alpha_1\alpha_2\alpha_3\alpha_4q_0 \,, \qquad \mathcal E\equiv\sum_I D_I \,.
    \label{eq:app-charge-phase}
\end{equation}
Taking the absolute value therefore gives
\begin{equation}
    \mathcal E \leq N+|q_{4'}|+|q_4|+|q_0| \,.
    \label{eq:app-energy-upper}
\end{equation}

On the other hand, for real $x,y$,
\begin{equation}
    \sqrt{(1+x^2)(1+y^2)} \geq 1+|xy| \,, 
\end{equation}
with equality if and only if $|x|=|y|$. Applying this inequality to the pairs $(f_1,f_2)$ and $(f_3,f_4)$, and then summing over the branes, gives
\begin{equation}
    \begin{aligned}
        \mathcal E &\geq \sum_I \left(1+\left|f_1^{(I)}f_2^{(I)}\right|\right) \left(1+\left|f_3^{(I)}f_4^{(I)}\right|\right) \\
        &\geq N+|q_{4'}|+|q_4|+|q_0| \,.
    \end{aligned}
    \label{eq:app-energy-lower}
\end{equation}
The upper and lower bounds coincide, so every inequality must be saturated.

The first inequality in \eqref{eq:app-energy-lower} then requires
\begin{equation}
    |f_1^{(I)}|=|f_2^{(I)}| \,, \qquad |f_3^{(I)}|=|f_4^{(I)}|
\end{equation}
on every brane. Saturation of the triangle inequalities fixes the relative signs as well. In particular, since the $N$ term is positive, all nonzero terms on the right-hand side of \eqref{eq:app-charge-phase} must have the same positive sign. Thus $e^{ic}=1$, and
\begin{equation}
    f_2^{(I)}=\eta_1f_1^{(I)} \,, \qquad f_4^{(I)}=\eta_2f_3^{(I)}\,, \qquad \eta_1=-\alpha_1\alpha_2 \,, \quad \eta_2=-\alpha_3\alpha_4 \,.
\label{eq:signcon}
\end{equation}
The same $\eta_1,\eta_2$ apply to every brane. Since the angles are defined by the principal arctangent, they obey the same pairwise relations. This proves \eqref{eq:thetapairing}.

Using $J_1J_2=\Gamma^{1234}$ and $J_3J_4=\Gamma^{5678}$, the relations \eqref{eq:signcon} give
\begin{equation}
    \Gamma^{1234}\epsilon=\eta_1\epsilon \,, \qquad \Gamma^{5678}\epsilon=\eta_2\epsilon \,.
\end{equation}
These imply
\begin{equation}
    (J_1+\eta_1J_2)\epsilon=0 \,, \qquad (J_3+\eta_2J_4)\epsilon=0 \,,
\end{equation}
and hence $X_I\epsilon=0$ for every brane. Since $P_8$ anticommutes with $X_I$, the original $\kappa$-symmetry conditions therefore reduce to
\begin{equation}
    P_8\epsilon=-\epsilon \,.
\end{equation}
Together, these are the three projections in \eqref{eq:threeproj}, leaving four supercharges.

\bibliographystyle{JHEP}
\bibliography{branes.bib}

\end{document}